\documentclass[10pt,fleqn]{article}
\usepackage[utf8]{inputenc}
\usepackage[T1]{fontenc}
\usepackage[top=1in,bottom=1in,left=1.2in,right=1.2in]{geometry}
\usepackage{authblk}
\usepackage{amsmath,amsfonts,amssymb}

\numberwithin{equation}{section}

\newcommand{\ket}[1]{\rvert#1\rangle}

\newcommand{\re}[1]{\mathfrak{Re}\left(#1\right)}
\newcommand{\im}[1]{\mathfrak{Im}(#1)}

\title{Generating function for the Bannai-Ito polynomials}

\author[1]{Geoffroy Bergeron}

\author[1]{Luc Vinet}

\author[2]{Satoshi Tsujimoto}

\affil[1]{Centre de recherches math\'ematiques, Universit\'e de Montr\'eal, P.O. Box 6128, Centre-ville Station, Montr\'eal, Canada H3C 3J7}
\affil[2]{Department of Applied Mathematics and Physics, Graduate School of Informatics, Kyoto University, Kyoto, 606-8501, Japan}

\date{\today}

\begin{document}

\maketitle
\thispagestyle{empty}
\hrule
\begin{abstract}\noindent
A generating function for the Bannai-Ito polynomials is derived using the fact that these polynomials are known to be essentially the Racah or $6j$ coefficients of the $\mathfrak{osp}(1|2)$ Lie superalgebra. The derivation is carried in a realization of the recoupling problem in terms of three Dunkl oscillators.
\end{abstract} 
\hrule

\section{Introduction}
In a previous paper \cite{Bergeron2015}, generating functions for the dual -1 Hahn polynomials were derived using the Clebsch-Gordan problem of the $\mathfrak{osp}(1|2)$ Lie superalgebra. In the present case, we exploit again the fact that $\mathfrak{osp}(1|2)$ is the dynamical algebra of a parabosonic or Dunkl oscillator. The generating function of the Bannai-Ito polynomials is found by using the wavefunctions of this system and recalling \cite{BIasRacah} that the Racah coefficients for $\mathfrak{osp}(1|2)$ are given in terms of these polynomials.

Related approaches using wavefunction realizations of dynamical algebra to derive identities for orthogonal polynomials have been presented previously \cite{Zhedanov1993,Koelink1996,VanderJeugt1997a,VanderJeugt1998}. In particular, \cite{Genest2014a} uses manipulations of wavefunctions similar to the ones that will be presented here to derive an integral representation of recoupling coefficients.

\subsection{The Bannai-Ito polynomials}

The Bannai-Ito polynomials, introduced in \cite{Bannai1984}, denoted here by $B_n(x)$, depend on four parameters $\{r_1,r_2,\rho_1,\rho_2\}$ and can be defined \cite{Tsujimoto2012}, see also \cite{Vinet2010}, as the functions diagonalizing the difference operator
\begin{align*}
\frac{(x-\rho_1)(x-\rho_2)}{2x}(I-P_x)+\frac{(x-r_1+1/2)(x-r_2+1/2)}{2x+1}(P_x D_x - I),
\end{align*}
where $P_x$ is the reflection operator acting on functions of $x$ as $P_x f(x) = f(-x)$ and $D_x$ is the forward shift operator acting as $D_x f(x) = f(x-1)$ and with the eigenvalues $\lambda_n$ given by
\begin{equation*}
\lambda_n =
\begin{cases}
\displaystyle \frac{n}{2} &\text{ for } n \text{ even,}\\
\displaystyle r_1+r_2-\rho_1-\rho_{2}-\frac{n+1}{2} &\text{ for } n \text{ odd.}
\end{cases}
\end{equation*}
They satisfy a three-term recurrence relation
\begin{align*}
x B_n(x) = B_{n+1}(x) + (\rho_1-a_n-c_n)B_n(x) + a_{n-1}c_n B_{n-1}(x),
\end{align*}
with coefficients
\begin{equation}\label{recuAn}
a_n =
\begin{cases}
\displaystyle \frac{(n+2\rho_1-2 r_1+1)(n+2 \rho_1-2 r_2+1)}{4(n+\rho_1+\rho_2-r_1-r_2+1)} &\text{for }n\text{ even,}\\
\\
\displaystyle \frac{(n + 2 \rho_1 + 2 \rho_2 - 2 r_1 - 2 r_2 + 1)(n + 2 \rho_1 + 2  \rho_2 + 1)}{4 (n + \rho_1 + \rho_2 - r_1 - r_2 + 1)} &\text{for }n\text{ odd,}
\end{cases}
\end{equation}
\begin{equation}\label{recuBn}
c_n = 
\begin{cases}
\displaystyle\frac{-n (n - 2 r_1 - 2 r_2)}{4 (n + \rho_1 + \rho_2 - r_1 - r_2)} &\text{for }n\text{ even,}\\
\\
\displaystyle\frac{-(n + 2 \rho_2 - 2 r_2) (n + 2 \rho_2 - 2 r_1)}{4 (n + \rho_1 + \rho_2 - r_1 - r_2)} &\text{for }n\text{ odd,}
\end{cases}
\end{equation}
and initial conditions $B_{-1}(x) = 0$, $B_0(x)=1$. The possible choices of truncation conditions for the recurrence relation are, for $N$ even,
\begin{equation}
 2(r_i - \rho_k)=N+1, \quad i,k=1,2,
\end{equation}
and, for $N$ odd,
\begin{equation}
 \rho_1 + \rho_2 = -(N+1)/2, \quad \text{ or } \quad r_1 + r_2 = (N+1)/2.
\end{equation}
In this work, the truncation conditions used are
\begin{equation}
 2(r_2 - \rho_1) = N+1,\, \text{ for }N\text{ even, }\rho_1 + \rho_2 = -(N+1)/2,\,\text{ for }N\text{ odd.}
\end{equation}
The orthogonality of the Bannai-Ito polynomials
\begin{equation}
\sum\limits_{S=0}^N w_S B_n(x_S) B_m(x_S) = h_N \delta_{nm}, \quad S = 0,...\,,N,
\end{equation}
is with respect to a discrete measure of weights $w_S$ on the grid $x_S$, for $S = 2s+p \in\{0,...,N\}$ and $p\in\{0,1\}$, with normalization $h_N$  where
\begin{equation}\label{ws}
w_S = \frac{(-1)^p (\rho_1-r_1+1/2)_{s+p}(\rho_1-r_2+1/2)_{s+p}(\rho_1+\rho_2+1)_{s}}{(\rho_1+r_1+1/2)_{s+p}(\rho_1+r_2+1/2)_{s+p}(1)_{s}(\rho_1-\rho_2+1)_{s}},
\end{equation}
with $(a)_m=a(a+1)...(a+m-1)$ the rising Pochhammer symbol, and
\begin{align}
x_S &= \frac{(-1)^S (S +2 \rho_1 +1/2) - 1/2}{2},\\
h_N &= \label{hn}
\begin{cases}
\displaystyle\frac{(2\rho_1+1)_{N/2}(r_1-\rho_2+1/2)_{N/2}}{(\rho_1-\rho_2+1)_{N/2}(\rho_1+r_1+1/2)_{N/2}}, &N\text{ even,}\\
\displaystyle\frac{(2\rho_1+1)_{(N+1)/2}(r_1+r_2)_{(N+1)/2}}{(\rho_1+r_1+1/2)_{(N+1)/2}(\rho_1+r_2+1/2)_{(N+1)/2}}, &N\text{ odd,}
\end{cases}
\end{align}
where $N = |\rho_2+r_2|+r_2-\rho_2-2\rho_1-1$.

\subsection{The $\mathfrak{osp}(1|2)$ algebra}
The $\mathfrak{osp}(1|2)$ algebra is generated by two odd elements $K_\pm$ and one even element $K_0$, relative to a $\mathbb{Z}_2$-grading. The presentation used in this paper makes this grading explicit by the introduction of a grade involution operator $R$ that commutes/anticommutes with the even/odd elements of the algebra. This presentation, also referred to as the $\mathfrak{sl}_{-1}(2)$ algebra \cite{Tsujimoto2011} in the literature, is given by the four generators $K_0$, $K_\pm$ and $R$ together with the relations
\begin{align}\label{osp12}
[ K_0,K_\pm] = \pm K_\pm, \quad [K_0,R]=0, \quad \{K_+,K_-\}=2 K_0, \quad \{K_\pm,R\}=0, \quad R^2=1,
\end{align}
with $[a,b]=ab-ba$ and $\{a,b\}=ab+ba$. The Casimir operator for the algebra as presented in \eqref{osp12} is given by
\begin{align}\label{casimir}
 C &= (K_+K_- - K_0 + 1/2)R.
\end{align}

The irreducible positive-discrete series representations of $\mathfrak{osp}(1|2)$ are then labeled by two numbers $(\mu,\epsilon)$ where $\mu\geq 0$ and $\epsilon=\pm1$. The actions of the generators on the orthonormal basis vectors $|n,\mu,\epsilon\rangle$ with $n \in \mathbb{N}$ are
\begin{alignat}{2}
  \label{basis}
\begin{aligned}
 K_{0}\,\ket{n,\mu,\epsilon}&= (n + \mu + 1/2)\,\ket{n,\mu,\epsilon},&\quad R\ket{n,\mu,\epsilon} &= \epsilon\, (-1)^n\,\ket{n,\mu,\epsilon},
 \\
 K_{+}\,\ket{n,\mu,\epsilon} &= \sqrt{[n+1]_\mu}\,\ket{n+1,\mu,\epsilon},&\quad K_{-}\,\ket{n,\mu,\epsilon} &= \sqrt{[n]_\mu}\,\ket{n-1,\mu,\epsilon},
 \end{aligned}
\end{alignat}
where $[n]_\mu = n + \mu(1-(-1)^n)$. In these representations, the Casimir \eqref{casimir} assumes the value
\begin{align*}
 C|n,\mu,\epsilon\rangle = - \epsilon \mu |n,\mu,\epsilon\rangle.
\end{align*}

\subsection{Realization as a dynamical algebra}
The presentation \eqref{osp12} of $\mathfrak{osp}(1|2)$ can be realized \cite{BIalgebraS2,Dunkl3D} in terms of operators acting on functions of a real variable $x$. Let $P_x$ denote the parity operator acting on functions as $P_x f(x) = f(-x)$. The $\mathbb{Z}_2$-Dunkl derivative is defined by
\begin{align*}
\mathfrak{D}_x = \partial_x + \frac{\mu}{x} (1-P_x).
\end{align*}
The $\mathfrak{osp}(1|2)$ algebra is realized under the following identification of the generators:
\begin{align}\label{realization}
K_0 = -\frac{1}{2}\mathfrak{D}_x^2 + \frac{1}{2}x^2, \quad K_\pm = \frac{1}{\sqrt{2}}(x\mp\mathfrak{D}_x), \quad R=P_x.
\end{align}
This casts $\mathfrak{osp}(1|2)$ as the dynamical algebra of the parabose oscillator \cite{Tsujimoto2011} whose Hamiltonian $H$ is the operator that realizes $K_0$. It follows that the position operator and its associated eigenvectors are
\begin{align}\label{Xoperator}
X = \frac{1}{\sqrt{2}}(K_++K_-), \quad X|x,\mu,\epsilon\rangle = x |x,\mu,\epsilon\rangle.
\end{align}
The representation basis \eqref{basis} corresponds to the energy eigenstates with eigenvalues $E = n + \mu +1/2$ and can be modeled by the wavefunctions $\Psi_n^{\mu,\epsilon}(x)$ defined through
\begin{align}\label{1Dwavefunctions}
\Psi_n^{\mu,\epsilon}(x) &= \langle x, \mu, \epsilon | n, \mu, \epsilon \rangle.
\end{align}

\subsection{The Racah problem of $\mathfrak{osp}(1|2)$}
The $\mathfrak{osp}(1|2)$ algebra also forms a Hopf algebra \cite{LaplaceDunklS2} where the coproduct $\Delta$ is given in the presentation \eqref{osp12} as
\begin{equation}\label{coproduct}
 \Delta(K_0) = K_0\otimes1 + 1\otimes K_0, \quad \Delta(R) = R\otimes R,\quad \Delta(K_\pm) = K_\pm\otimes R + 1\otimes K_\pm.
\end{equation}
This Hopf algebra structure induces an action of $\mathfrak{osp}(1|2)$ on tensor products of modules. Consider the following threefold tensor product of irreducible representations
\begin{align}\label{threefold}
(\mu_1,\epsilon_1)\otimes(\mu_2,\epsilon_2)\otimes(\mu_3,\epsilon_3).
\end{align}
One can decompose this product of representations in a direct sum of irreducible representations in two different ways, corresponding to the order in which the coproduct is used to induce an action of $\mathfrak{osp}(1|2)$ on \eqref{threefold}, either $\Delta\otimes 1\circ\Delta$ or  $1\otimes\Delta\circ\Delta$. Both cases specify an algebra homomorphism $\mathfrak{osp}(1|2) \rightarrow \mathfrak{osp}(1|2)\otimes\mathfrak{osp}(1|2)\otimes\mathfrak{osp}(1|2)$ and an associated decomposition of threefold tensor products of representations into direct sums of irreducible representations
\begin{equation}\label{threefoldmap}
\begin{split}
(\mu_1,\epsilon_1)\otimes(\mu_2,\epsilon_2)\otimes(\mu_3,\epsilon_3) & \overset{\Delta\otimes 1\circ\Delta}{\cong}
\bigoplus_{u} (\mu_{(12)3}(u),\epsilon_{(12)3}(u)),\\
& \overset{1\otimes \Delta\circ\Delta}{\cong} \bigoplus_{v}(\mu_{1(23)}(v),\epsilon_{1(23)}(v)).
\end{split}
\end{equation}
In fact, by the coassociativity of the coproduct, we have that two irreducible representations connected in such a way are isomorphic
\begin{align}\label{sameRep}
(\mu_{(12)3},\epsilon_{(12)3}) \cong (\mu_{1(23)},\epsilon_{1(23)})
\end{align}
and thus, we will only keep the notation distinguishing the two in the labels when relevant.

The basis constructed as in \eqref{basis} for these representations does not uniquely determine the map \eqref{threefoldmap} on the basis vectors themselves, but a canonical choice of supplementary labels exists that removes the degeneracy. One demands that the basis vectors of $(\mu_{(12)3},\epsilon_{(12)3})$, (respectively $(\mu_{1(23)},\epsilon_{1(23)})$), diagonalize the intermediate Casimir operator $C_{12} = \Delta(C)\otimes 1$, (resp. $C_{23} = 1\otimes\Delta(C)$). Thus, denoting the action of $\Delta\otimes 1\circ\Delta = 1\otimes \Delta\circ\Delta$ on generators $A \in \mathfrak{osp}(1|2)$ by
\begin{align}\label{hatDefinition}
\Delta\otimes 1\circ\Delta : A \mapsto \hat{A} \in \mathfrak{osp}(1|2)\otimes\mathfrak{osp}(1|2)\otimes\mathfrak{osp}(1|2),
\end{align}
and knowing the two modules $(\mu_{(12)3},\epsilon_{(12)3})$ and $(\mu_{1(23)},\epsilon_{1(23)})$ are identical as $\mathfrak{osp}(1|2)$ representations, we have that basis vectors for both satisfy 
\begin{alignat}{2}
\label{basisCoupled}
\begin{aligned}
 \hat{K}_{0}\,\ket{n_{123}}&= (n_{123} + \mu_{123} + 1/2)\,\ket{n_{123}},&\quad \hat{R}\,\ket{n_{123}} &=\epsilon_{123}\, (-1)^{n_{123}}\,\ket{n_{123}},
 \\
 \hat{K}_{+}\,\ket{n_{123}} &= \sqrt{[n_{123}+1]_{\mu_{123}}}\,\ket{n_{123}+1},&\quad \hat{K}_{-}\,\ket{n_{123}} &= \sqrt{[n_{123}]_{\mu_{123}}}\,\ket{n_{123}-1},
 \\
 \hat{C}\, \ket{n_{123}} &= -\mu_{123}\,\epsilon_{123}\, \ket{n_{123}}, &&
 \end{aligned}
\end{alignat}
where $\ket{n_{123}}$ stands for either $\ket{n_{(12)3},\mu_{(12)3},\epsilon_{(12)3}}$ or $\ket{n_{1(23)},\mu_{1(23)},\epsilon_{1(23)}}$. The degeneracy is then lifted through the actions of the intermediate Casimirs
\begin{subequations}
\begin{align}
 \label{basisCoupled1}
 C_{12}\, \ket{n_{(12)3},\mu_{(12)3},\epsilon_{(12)3}} &= -\mu_{12}\,\epsilon_{12}\,\ket{n_{(12)3},\mu_{(12)3},\epsilon_{(12)3}},\\
 \label{basisCoupled2} 
 C_{23}\, \ket{n_{1(23)},\mu_{1(23)},\epsilon_{1(23)}} &= -\mu_{23}\,\epsilon_{23}\,\ket{n_{1(23)},\mu_{1(23)},\epsilon_{1(23)}}.
\end{align}
\end{subequations}
These bases are not the same since $[C_{12},C_{23}] \neq 0$. The $\mathfrak{osp}(1|2)$ Racah problem consists in determining the overlaps $\mathcal{R}$ between the two bases \eqref{basisCoupled}
\begin{align}\label{RacahDef}
\mathcal{R} = \langle n_{(12)3},\mu_{(12)3},\epsilon_{(12)3} \ket{n_{1(23)},\mu_{1(23)},\epsilon_{1(23)}}.
\end{align}

\subsection{Outline}
We will first explain the realization of the Racah problem in terms of a system of three parabose harmonic oscillators and will indicate how this realization relates to generating functions in section 2. Section 3 gives the explicit expressions of the angular wavefunctions in each parity case of the parameters and a derivation of their asymptotic form in the relevant limits. Finally, section 4 contains the derivation of the generating functions and is followed by a brief conclusion.

\section{Realization of the Racah decomposition}

The Racah problem of $\mathfrak{osp}(1|2)$ can be expressed within the dynamical algebra realization by considering three uncoupled parabose oscillators in the Cartesian coordinates $\{x$, $y$, $z\}$. The total Hamiltonian for this system is simply the sum of the separate Hamiltonians
\begin{align*}
H_{xyz} = H_x + H_y + H_z = K_0\otimes1\otimes1 + 1\otimes K_0 \otimes 1 + 1\otimes 1\otimes K_0 = \hat{K}_0.
\end{align*}
The Schr\"odinger equation $H_{xyz} |\psi\rangle = E_{xyz} |\psi\rangle$ manifestly separates in the Cartesian coordinates. In \cite{Dunkl3D}, it was shown that it also separates in spherical coordinates. This separation is associated to the symmetries generated by the intermediate Casimir operators $C_{12}$ and $C_{23}$. In fact, the spherical wavefunctions are constructed \cite{LaplaceDunklS2} using the basis \eqref{basisCoupled}.  Not surprisingly then, the Racah problem is directly related to the different possible choices in the construction of the spherical coordinates.

\subsection{Spherical coordinates realization}

The position operator $X$ introduced in \eqref{Xoperator} can naturally be extended to a set of three operators acting on threefold tensor product of irreducible representations as
\begin{align*}
X = X\otimes 1 \otimes 1, \quad Y = 1\otimes X\otimes 1, \quad Z = 1\otimes 1\otimes X,
\end{align*}
where the $X$ operator in the right-hand side is the one defined in \eqref{Xoperator}. From these, one can define the radial operator $X^2+Y^2+Z^2$. It commutes \cite{BIalgebraS2} with the intermediate Casimirs $C_{12}$ and $C_{23}$. Thus, the two bases introduced in \eqref{basisCoupled} do not differ in their radial parts and the Racah problem is entirely determined by the angular wavefunctions. We may as well take the radius to be fixed and consider the Racah problem on a fixed eigenspace of the radial operator $\hat{X}^2$.

The angular wavefunctions will be defined as the functions satisfying \eqref{basisCoupled1} or \eqref{basisCoupled2} under the action of the $\mathfrak{osp}(1|2)$ algebra in the coordinate realization and under the constraint $x^2+y^2+z^2 = 1$, where $x$, $y$ and $z$ are the eigenvalues of the $X$, $Y$ and $Z$ operators, respectively. As such, these functions are defined on the two-dimensional sphere and can be parametrized by two angles $\theta$ and $\phi$. We choose these angles to be related to the Cartesian coordinates as usual through
\begin{align}\label{sphericalCoord}
x = \sin\theta \cos\phi, \quad y = \sin\theta \sin\phi, \quad z = \cos\theta.
\end{align}
Using these relations, the realization \eqref{realization} of $\mathfrak{osp}(1|2)$ can be expressed as differential operators in the angular coordinates \cite{Dunkl3D}. The angular wavefunctions are then given by
\begin{align*}
\mathcal{Y}_{n_{(12)3}}^{\mu_{(12)3},\epsilon_{(12)3}}(\theta,\phi) = \langle \theta,\phi \, \ket{n_{(12)3},\mu_{(12)3},\epsilon_{(12)3}}, \quad\text{with}\quad x^2+y^2+z^2=1.
\end{align*}
A similar expression is defined for the other basis with a different set of angular variables $\{ \alpha,\beta \}$ by
\begin{align*}
\mathcal{Z}_{n_{1(23)}}^{\mu_{1(23)},\epsilon_{1(23)}}(\alpha,\beta) = \langle \alpha,\beta \, \ket{n_{1(23)},\mu_{1(23)},\epsilon_{1(23)}}, \quad\text{with}\quad x^2+y^2+z^2=1.
\end{align*}
It is possible to relate the second set of variables to the first by observing that a permutation of the terms in the threefold tensor product of irreducible representations \eqref{threefold} maps the basis \eqref{basisCoupled1} to \eqref{basisCoupled2}. Explicitly, this permutation is the cycle $(\,1\,2\,3\,)$ acting on the Cartesian coordinates $\{x,y,z\}$. In terms of the angular variables, this corresponds to the relations
\begin{align}\label{newVariables}
\sin\alpha\cos\beta = \sin\theta\sin\phi, \quad \sin\alpha\sin\beta = \cos\theta, \quad \cos\alpha = \sin\theta\cos\phi.
\end{align}
In view of \eqref{sameRep}, the decomposition of these angular wavefunctions onto each other exists and will have the Racah coefficients as overlaps
\begin{align}\label{rawDecomposition}
\mathcal{Z}_{n_{1(23)}}^{\mu_{1(23)},\epsilon_{1(23)}}(\alpha(\theta,\phi),\beta(\theta,\phi)) = \sum \mathcal{R}\, \mathcal{Y}_{n_{(12)3}}^{\mu_{(12)3},\epsilon_{(12)3}}(\theta,\phi),
\end{align}
where $\alpha(\theta,\phi)$ and $\beta(\theta,\phi)$ are obtained from \eqref{newVariables}.

\subsection{Exact form of the decomposition}

Let us now make details explicit. First consider a basis vector of \eqref{threefold}, which we here denote by \newline$\ket{n_1,\mu_1,\epsilon_1} \otimes \ket{n_2,\mu_2,\epsilon_2} \otimes \ket{n_3,\mu_3,\epsilon_3}$. In view of \eqref{threefoldmap}, we may write a decomposition of the form
\begin{align}\label{threefoldDecomp}
\ket{n_1,\mu_1,\epsilon_1}\otimes\ket{n_2,\mu_2,\epsilon_2}\otimes\ket{n_3,\mu_3,\epsilon_3} = \sum_u\,\mathcal{C}_u\, \ket{n_{123},\mu_{123},\epsilon_{123}}_u.
\end{align}
For this equality to hold given the action of $\hat{K}_0$ and $\hat{R}$ and knowing that the representation parameters $\mu_{123}$ and $\epsilon_{123}$ cannot depend on the basis label $n_{123}$ one obtains the following relations
\begin{align}\label{globalParameters}
&n_{123} = n_1 + n_2 + n_3 - N, \quad \mu_{123} = \mu_1 + \mu_2 + \mu_3 + 1 + N, \quad \epsilon_{123} = \epsilon_1\epsilon_2\epsilon_3 (-1)^N,
\end{align}
where $N \in [0,n_1+n_2+n_3] \subset \mathbb{N}$. The difference between the two bases \eqref{basisCoupled1} and \eqref{basisCoupled2} arises when considering the operators $C_{12}$ and $C_{23}$. Being intermediate Casimirs, these operators satisfy
\begin{equation*}
[C_{12},\Delta(A)\otimes1]=0=[C_{23},1\otimes\Delta(A)] \quad \forall \quad A \in \mathfrak{osp}(1|2).
\end{equation*}
Demanding their diagonalisation as in \eqref{basisCoupled1} or \eqref{basisCoupled2} requires the decomposition \eqref{threefoldDecomp} to solve the Clebsch-Gordan problem \cite{Bergeron2015}, if one focuses only on the first or second pair of terms in the tensor product \eqref{threefold}. It is known that the parameters involved in the Clebsch-Gordan decomposition into the product representation $(\mu_i,\epsilon_i)\otimes(\mu_j,\epsilon_j)$ must verify, for $n_i + n_j \geq q \in \mathbb{N}$,
\begin{align}\label{clebschParameters}
n_{ij} = n_i + n_j - q,\quad \mu_{ij} = \mu_i + \mu_j + q + 1/2,\quad \epsilon_{ij} = \epsilon_i\epsilon_j(-1)^q,
\end{align}
corresponding to the diagonalization of the intermediate Casimir $C_{ij}$.
Rewriting \eqref{globalParameters} in view of \eqref{clebschParameters}, one has that the labels of basis vectors in \eqref{threefoldDecomp} with non-vanishing overlap and where $\ket{n_{123},\mu_{123},\epsilon_{123}}$ diagonalizes $C_{ij}$ are related by the following equations
\begin{align}\label{parameters1}
n_{123} &= n_{ij} + n_k - l, & \mu_{123} &= \mu_{ij} + \mu_k + 1/2 + l, & \epsilon_{123} &= \epsilon_{ij}\epsilon_k(-1)^{l},\\
l &= N - q, & N &\in [0,n_1+n_2+n_3] \subset \mathbb{N}, &\implies q &\in [0,N]  \subset \mathbb{N}. \label{parameters3}
\end{align}
where $i,j \in \{(1,2),(2,3)\}$ and $i,j \neq k \in \{1,2,3\}$ index the terms of the threefold tensor product \eqref{threefold}.

For given values of $\mu_1$, $\mu_2$, $\mu_3$, $\epsilon_1$, $\epsilon_2$, $\epsilon_3$ and $N$, the parameters $\mu_{123}$ and $\epsilon_{123}$ are fixed and, since $l \ge 0$, there are $N+1$ ways of choosing $q$. Thus, the decomposition of the tensor product of three irreducible $\mathfrak{osp}(1|2)$ representations can be expressed as
\begin{align*}
(\mu_1,\epsilon_1)\otimes(\mu_2,\epsilon_2)\otimes(\mu_3,\epsilon_3) \cong \bigoplus_{N=0}^\infty \bigoplus_{q=0}^{N}(\mu_{123}(N),\epsilon_{123}(N))_q,
\end{align*}
where $q$ indexes as in \eqref{clebschParameters} the possible eigenvalues of the intermediate Casimir $C_{ij}$.

Consider now the Racah coefficients $\mathcal{R}$ as given in \eqref{RacahDef} where both basis vectors come from one of the two different decompositions \eqref{threefoldmap} of the same threefold tensor product of irreducible representations \eqref{threefold}. As the two modules in consideration are identical as $\mathfrak{osp}(1\rvert 2)$ modules, the Racah coefficients vanish if the labels of the basis vectors differ. The only free parameter in non-zero coefficients is the value of the intermediate Casimirs. Thus, writing as $K$ and $S$ those free parameters indexing the values of the intermediate Casimirs for the two basis vectors in the overlap, the Racah decomposition will explicitly be written as
\begin{align*}
\ket{n_{123},\mu_{123},\epsilon_{123},\mu_{12}(S)} = \sum_{K = 0}^N \mathcal{R}_{S,K,N}^{\mu_1,\mu_2,\mu_3} \ket{n_{123},\mu_{123},\epsilon_{123},\mu_{23}(K)}.
\end{align*}
This equation can be rewritten in terms of the wavefunctions. As said, the overlap between two such wavefunctions is directly proportional to the Bannai-Ito polynomials \cite{LaplaceDunklS2}:
\begin{align}\label{ZYequation}
\mathcal{Z}_S^N(\alpha(\theta,\phi),\beta(\theta,\phi)) &= \sum\limits_{K=0}^N \mathcal{R}_{S,K,N}^{\mu_1,\mu_2,\mu_3} \mathcal{Y}_K^N(\theta,\phi),\\
\label{RacahWave}
\mathcal{R}_{S,K,N}^{\mu_1,\mu_2,\mu_3}&=\Phi_S^N\sqrt{\frac{w_{S}}{h_N u_1 u_2...u_K}}B_K(x_S;\rho_1,\rho_2,r_1,r_2)
\end{align}
with $w_S$, $x_S$ and $h_N$ as in \eqref{ws} and \eqref{hn} and where the $B_K$ are the Bannai-Ito polynomials. The $u_i$ are given by $u_i = a_{n-1}b_n$ with $a_n$ and $b_n$ as in \eqref{recuAn} and \eqref{recuBn}. The choice of phase $\Phi_S^N$ is different than in \cite{LaplaceDunklS2}. In this work, writing $N=2n+t\in\mathbb{N}$ and $S=2s+p$ with $p,t \in \{0,1\}$ the phase is given by
\begin{equation}
\Phi_S^N = (-1)^{n+t(1-p)}.
\end{equation}
The connection between the parameters of the threefold tensor product \eqref{threefold} and the parameters of the Bannai-Ito polynomials in \eqref{RacahWave} is as follows
\begin{equation}\label{RacahParam}
\begin{split}
\rho_1 &= \frac{\mu_2+\mu_3}{2}, \quad \rho_2=\frac{\mu_1+\mu}{2}, \quad r_1=\frac{\mu_3-\mu_2}{2}, \quad r_2=\frac{\mu-\mu_1}{2}\\
\mu &= (-1)^N(N+1+\mu_1+\mu_2+\mu_3).
\end{split}
\end{equation}

\subsection{Generating function from the Racah problem}

The wavefunction realization of the Racah decomposition \eqref{ZYequation} leads to a functional decomposition with coefficients proportional to the Bannai-Ito polynomials \cite{BIasRacah}. To obtain generating functions, one needs to reduce the right-hand side of \eqref{ZYequation} to a power series of a single variable. As shall be explicit, the angular wavefunctions are polynomials of trigonometric functions which reduces, under some asymptotic expansion, to their leading terms. Monomials are obtained from the expansion by the simultaneous introduction of a suitable relation between the angle variables. However, this procedure must be carried while preventing the trivialization of the left-hand side of \eqref{ZYequation}.

In view of the form of the wavefunctions given in section 3.1, one is led to consider the expansion $\rvert\theta\rvert \rightarrow 0$. To prevent a trivialization we introduce, as follows, the finite variable $z=\cos\alpha$ and use \eqref{newVariables} under the asymptotic expansion to obtain the following  
\begin{align} \label{zRelations}
\sin\alpha = \sqrt{1-z^2}, \quad \sin\beta = \frac{1}{\sqrt{1-z^2}}, \quad \cos\beta = i\frac{z}{\sqrt{1-z^2}},
\end{align}
The finiteness of $z$ together with \eqref{newVariables} implies $\im{\phi} \rightarrow \infty$. With $\im{\phi} \geq 0$, one demands that the following limits be defined and give
\begin{align*}
\cosh \im{\phi} \sin\theta \rightarrow \lambda, \quad \sinh \im{\phi} \sin\theta \rightarrow \lambda,
\end{align*}
such that compatibility with \eqref{newVariables} and \eqref{zRelations} is maintained and
\begin{align}\label{zDef}
z = \lambda e^{- i \re{\phi}}.
\end{align}
Using \eqref{newVariables} and \eqref{zRelations}, the following useful relation can be obtained under the asymptotic limit
\begin{align}\label{sinTOcos}
\sin \phi \approx i \cos\phi.
\end{align}

Under this asymptotic limit, the decomposition \eqref{ZYequation} will take the form of a generating function for the sum of two Bannai-Ito polynomials
\begin{equation}\label{GenFunDecomp}
\mathcal{Z}_S^N(z) = \sum\limits_{K=0}^N \mathcal{R}_{S,K,N}^{\mu_1,\mu_2,\mu_3} \mathcal{Y}_K^N(z),
\end{equation}
where $\mathcal{Y}_K^N(z)$ is a sum of two monomials of the $z$ variable. It should be noted, in view of \eqref{zDef} and since the parameters $\lambda$ and $\re{\phi}$ are not fixed, that $z$ can be any complex number.

\section{Wavefunctions and their asymptotic forms}

As the Racah problem is fully contained in the overlaps of angular wavefunctions, one does not need a set of basis functions that reflects the full degeneracy of the Hamiltonian $H_{xyz}$. We shall use instead functions of definite parity on which the total Casimir is diagonal. This is justified by remembering that we have $\mathcal{R} \neq 0$ only when the overlap is between two basis vectors from the same eigenspace of the total Hamiltonian. These functions form a basis of the irreducible representations \eqref{sameRep} and are sufficient for our purpose but do not reflect the full degeneracy of the initial Shr\"odinger equation. This can be seen from the fact that the operators $R_i\,, \, i\in\{1,2,3\}$ commute with the total Hamiltonian, but not with the total Casimir, see \cite{LaplaceDunklS2}.

\subsection{Angular Wavefunctions}
The explicit form of the basis functions used in this work can be obtained by solving the relevant system of Dunkl differential equations. We assume $\epsilon_{123}=1$ for the rest of this work. In this case, the angular wavefunctions $\mathcal{Y}_K^N(\theta,\phi)$ for $K=0,...,N$ satisfy the following equations
\begin{align*}
\hat{C}\hat{R}\,\mathcal{Y}_K^N(\theta,\phi) &= -(N + \mu_1+\mu_2+\mu_3+1)\mathcal{Y}_K^N(\theta,\phi),\\
\hat{R}\,\mathcal{Y}_K^N(\theta,\phi) &= (-1)^N\mathcal{Y}_K^N(\theta,\phi),\\
C_{12} \,\mathcal{Y}_K^N(\theta,\phi) &= -(-1)^K (K+\mu_1+\mu_2)\mathcal{Y}_K^N(\theta,\phi),
\end{align*}
where these operators are defined on \eqref{threefold} using \eqref{hatDefinition} and the realization \eqref{realization}.

The solutions \cite{LaplaceDunklS2} correspond to (a subset of) the wavefunctions built on the basis \eqref{basisCoupled1} and are given, writing $N=2n+t$, $n \in \mathbb{N}$ and $K=2k+p \in \{0,...,N\}$ with $p,t \in \{0,1\}$, by
\begin{multline}\label{wave}
\mathcal{Y}_K^N(\theta,\phi) = A_{K} \Big\{ B_{K} \cos^{t}\theta \sin^{2k+2p}\theta P_{n-k-p}^{(2k+2p+\mu_1+\mu_2,\mu_3-1/2+t)}(\cos 2\theta) \mathcal{F}_K^+(\phi)\\
 + (-1)^{t} B_{K}^{-1} \cos^{1-t}\theta\sin^{2k+1}\theta P_{n-k-1}^{(2k+1+\mu_1+\mu_2,\mu_3+1/2-t)}(\cos 2\theta)\mathcal{F}_K^-(\phi) \Big\},
\end{multline}
where $A_{K}$ and $B_{K}$ are
\begin{align*}
A_K &= (-1)^{t K} \sqrt{\frac{(n-k+p(t-1))!\Gamma(n+k+\mu_1+\mu_2+\mu_3+3/2+pt)}{\Gamma(n+k+\mu_1+\mu_2+1+pt)\Gamma(n-k+\mu_3+1/2+p(t-1))}}, \\
B_K &= \left(\frac{n-k+\mu_3-1/2+t}{n+k+\mu_1+\mu_2+1}\right)^{(p-t)/2}.
\end{align*}
and where the $\mathcal{F}_K$ functions are as follows
\begin{align}\label{fkFunctionp}
&\begin{split}
 \mathcal{F}_K^+(\phi) &= \xi_K^+ \Big\{E_{K} P_{k+p}^{(\mu_2-1/2,\mu_1-1/2)}(\cos 2 \phi)\\
 & \qquad\qquad - (-1)^p E_{K}^{-1} \cos\phi\sin\phi P_{k+p-1}^{(\mu_2+1/2,\mu_1+1/2)}(\cos 2 \phi)\Big\},
\end{split}\\
\label{fkFunctionm}
&\begin{split}
\mathcal{F}_K^-(\phi) &= \xi_K^- \Big\{F_{K} \sin\phi P_{k}^{(\mu_2+1/2,\mu_1-1/2)}(\cos 2 \phi)\\
 & \qquad\qquad+ (-1)^p F_{K}^{-1} \cos\phi P_{k}^{(\mu_2-1/2,\mu_1+1/2)}(\cos 2 \phi)\Big\},
\end{split}
\end{align}
with
\begin{align*}
 \xi_K^+ &= \sqrt{\frac{(k+p)!\Gamma(k+\mu_1+\mu_2+1+p)}{2 \Gamma(k+\mu_1+1/2+p)\Gamma(k+\mu_2+1/2+p)}}, & E_{K} &= \left(\frac{k+1}{k+\mu_1+\mu_2+1}\right)^{p/2},\\
 \xi_K^- &= \sqrt{\frac{k!\Gamma(k+\mu_1+\mu_2+1)}{2 \Gamma(k+\mu_1+1/2)\Gamma(k+\mu_2+1/2)}}, & F_K &= \left(\frac{k+\mu_1+1/2}{k+\mu_2+1/2}\right)^{p/2}.
\end{align*}

A second wavefunction basis is obtained by reparametrizing the sphere in terms of the angular coordinates $\alpha,\beta$ as per \eqref{newVariables}. These wavefunctions, denoted $\mathcal{Z}_S^N(\alpha,\beta)$ for $S=0,...,N$, now satisfy the following equations
\begin{align*}
\hat{C}\hat{R}\,\mathcal{Z}_S^N(\alpha,\beta) &= -(N + \mu_1+\mu_2+\mu_3+1)\mathcal{Z}_S^N(\alpha,\beta),\\
\hat{R}\,\mathcal{Z}_S^N(\alpha,\beta) &= (-1)^N\mathcal{Z}_S^N(\alpha,\beta),\\
C_{23}\,\mathcal{Z}_S^N(\alpha,\beta) &= -(-1)^S (S+\mu_2+\mu_3)\mathcal{Z}_S^N(\alpha,\beta)
\end{align*}
and realize the basis defined by \eqref{basisCoupled2}. They can be written \cite{LaplaceDunklS2} in terms of the first basis of wavefunctions $\mathcal{Y}_K^N$ as
\begin{equation} \label{ZDef}
 \mathcal{Z}_S^N(\alpha,\beta) =
\begin{cases}
 (\,1\,2\,3\,)\, \mathcal{Y}_K^N(\pi-\alpha,\beta), &\text{for }N\text{ even,}\\
 (\,1\,2\,3\,)\, \mathcal{Y}_K^N(\alpha,\beta), &\text{for }N\text{ odd}.
\end{cases}
\end{equation}
where $(\,1\,2\,3\,)$ is the permutation cycle acting on the parameters $(\mu_1,\mu_2,\mu_3)$. This follows from the fact that this permutation induces a mapping from the basis of $(\mu_{(12)3},\epsilon_{(12)3})$ to the basis of $(\mu_{1(23)},\epsilon_{1(23)})$ when acting on the terms of the threefold tensor product \eqref{threefold}.

\subsection{Asymptotic Expansion}

We now derive the asymptotic expansion introduced in section 2.3 of the angular wavefunctions \eqref{wave}. One will need the leading term and the value at $1$ of Jacobi polynomials given by
\begin{equation*}
 P_n^{(a,b)}(x) \rightarrow 2^{-n}\binom{2n+a+b}{n}x^n, \qquad P_n^{(a,b)}(1) = \binom{n+a}{n}.
\end{equation*}
The polynomials in the $\mathcal{F}_K$ functions \eqref{fkFunctionp}, \eqref{fkFunctionm} have for variable $\cos 2 \phi \rightarrow \infty$ and only their leading terms will remain. Using \eqref{sinTOcos} in \eqref{fkFunctionp} or \eqref{fkFunctionm} while considering only the leading term leads to
\begin{align}\label{asFkp}
 \mathcal{F}_K^+(\phi) &\rightarrow \xi_K^+ E_K \binom{2k+2p+\mu_2+\mu_1-1}{k+p} \Psi_+ \cos^{2k+2p}\phi,\\
 \label{asFkm}\mathcal{F}_K^-(\phi) &\rightarrow \xi_K^- F_K \binom{2k+\mu_1+\mu_2}{k} \Psi_- \cos^{2k+1}\phi,
\end{align}
with $\Psi_+$ and $\Psi_-$ given by
\begin{align}\label{psiDef}
 \Psi_\pm =
\begin{cases}
\left[1-i(-1)^p \frac{k+p}{k+p+\mu_1+\mu_2}\left(\frac{k+\mu_1+\mu_2+1}{k+1}\right)^p\right]\quad &\text{for the case }+,\\
\\
\left[i + (-1)^p \left( \frac{k+\mu_2+1/2}{k+\mu_1+1/2}\right)^p \right]\quad &\text{for the case }-.
\end{cases}
\end{align}
Consider now the full wavefunctions \eqref{wave} under the asymptotic expansion. The remaining Jacobi polynomials are evaluated at $1$ as their arguments $\cos 2 \theta \rightarrow 1$. The remaining cosine terms also simply become $\cos \theta =1$. The sine terms approach zero, but will be compensated by the $\mathcal{F}_K$ functions which are divergent under the asymptotic expansion. Thus, leaving the sine terms, one is led to the following expressions for the asymptotic wavefunctions for $N=2n+t$
\begin{multline}\label{partialAsYeven}
 \mathcal{Y}_K^N(\theta,\phi) \rightarrow A_K \bigg\{ B_K \binom{n+k+p+\mu_1+\mu_2}{n-k-p} \mathcal{F}_K^+(\phi)\sin^{2k+2p}\theta\\
+ (-1)^{t} B_K^{-1} \binom{n+t+k+\mu_1+\mu_2}{n-k-1+t} \mathcal{F}_K^-(\phi)\sin^{2k+1}\theta\bigg\}.
\end{multline}
We now remind the reader that from \eqref{newVariables} and \eqref{zRelations} we have $\cos\phi\sin\theta = z$. By construction, this $z$ variable remains finite in the asymptotic expansion. Using \eqref{asFkp} and \eqref{asFkm} to rewrite \eqref{partialAsYeven} in terms of the $z$ variable leads to, for $N=2n+t$
\begin{multline}\label{fullAsYeven}
 \mathcal{Y}_K^N(z) = A_K \bigg\{\\
 \xi_K^+ B_K E_K \binom{n+k+p+\mu_1+\mu_2}{n-k-p}\binom{2k+2p+\mu_2+\mu_1-1}{k+p} \Psi_+ z^{2k+2p}\\
+(-1)^{t} \xi_K^- B_K^{-1} F_K \binom{n+t+k+\mu_1+\mu_2}{n+t-k-1}\binom{2k+\mu_1+\mu_2}{k} \Psi_- z^{2k+1}\bigg\}.
\end{multline}

\section{Generating functions}

In this section, we derive the main result, that is, the generating functions for the Bannai-Ito orthogonal polynomials. The wavefunctions in their asymptotic form being the sum of two monomials have not quite been brought to a monomial form. Thus, two degrees of the Bannai-Ito polynomials will appear in the coefficient of each power of the $z$ variable. This will not yield a proper generating function. However, once this intermediate result is obtained, it proves possible to disentangle the resulting power series with a trick involving analysis of the complex phase of each term. The next two subsections illustrate how the proper generating functions can be found using this two-step approach.

\subsection{Intermediate result}
The asymptotic expansion given in section 2.3 is constructed so that the trigonometric functions of the $\alpha$ and $\beta$ variables remain finite\footnote{Omitting the two poles $\{z=1,z=-1\}$ of the trigonometric functions of $\beta$.}. Thus, there is no expansion to be made on the left-hand side of \eqref{ZYequation} as defined in \eqref{ZDef} to obtain \eqref{GenFunDecomp}. One only needs to rewrite the functions in terms of the new variable $z$ through the use of \eqref{zRelations}. Using standard trigonometric relations for double angles and \eqref{zRelations}, we have
\begin{align*}
 \cos 2 \alpha = 2z^2 - 1,\qquad \cos 2 \beta = \frac{z^2+1}{z^2-1}.
\end{align*}

The functions of $\beta(z)$ in $\mathcal{Z}_S^N(z)$, amounting to the $\mathcal{F}_K$ functions in \eqref{fkFunctionp} and \eqref{fkFunctionm}, now depend on the parameter $S = 2s + p \in \{0,...,N\}$, $p \in \{0,1\}$ and have their parameters permuted by $(\,1\,2\,3\,)$ acting on $\{\mu_1,\mu_2,\mu_3\}$. These functions are expressed in terms of the new variable $z$ as
\begin{multline}
 \mathcal{F}_S^+(z) = \xi_S^+\Bigg[E_S P_{s+p}^{(\mu_3-1/2,\mu_2-1/2)}\left(\frac{z^2+1}{z^2-1}\right)\\
 - \frac{i z}{1-z^2} (-1)^p E_S^{-1} P_{s+p-1}^{(\mu_3+1/2,\mu_2+1/2)}\left(\frac{z^2+1}{z^2-1}\right)\Bigg],
\end{multline}
\begin{multline}
 \mathcal{F}_S^-(z) = \frac{\xi_S^-}{\sqrt{1-z^2}} \Bigg[F_S P_{s}^{(\mu_3+1/2,\mu_2-1/2)}\left(\frac{z^2+1}{z^2-1}\right)\\
  + i z (-1)^p F_S^{-1} P_{s}^{(\mu_3-1/2,\mu_2+1/2)}\left(\frac{z^2+1}{z^2-1}\right) \Bigg].
\end{multline}

Similarly, permuting the parameters, the angular wavefunctions \eqref{ZDef} are expressed in terms of $z$ when $N=2n+t$, $n \in \mathbb{N}$ and $t\in \{0,1\}$ as
\begin{multline}\label{fullGenFunEven}
  \mathcal{Z}_S^N(z) = A_S \bigg[ z^{t} B_S P_{n-s-p}^{(2s+2p+\mu_2+\mu_3,\mu_1-1/2+t)}(2z^2-1) \mathcal{F}_S^+(z)(1-z^2)^{s+p}\\
- z^{1-t} B_S^{-1} P_{n+t-s-1}^{(2s+1+\mu_2+\mu_3,\mu_1+1/2-t)}(2z^2-1)\mathcal{F}_S^-(z)(1-z^2)^{s+1/2}\bigg],
\end{multline}
where one must not forget to introduce the reflection in the $\alpha$ coordinate when $N$ is even.


\subsection{Proper generating functions}
We now turn to the problem of disentangling the quasi generating functions \eqref{fullGenFunEven}. Assuming $z$ to be a real variable, by observing \eqref{fullAsYeven} and \eqref{psiDef}, one can note that the phase information of the asymptotic wavefunctions $\mathcal{Y}_K^N(z)$ is given by $\Psi_+$ for even powers of $z$ and $\Psi_-$ for odd powers of $z$.

To disentangle the generating functions, we want to keep only the powers of $z$ coming from values of $K$ of the same parity. Thus, we only want to keep the terms with $p=0$ for the even powers of $z$ and the terms with $p=1$ for the odd powers of $z$. In this case, the matched $\Psi$ terms become
\begin{align}\label{phaseA}
 \Psi =
\begin{cases}
\left[1-i\frac{k}{k+\mu_1+\mu_2}\right]\quad &\text{for even powers of }z,\\
\\
\left[i - \frac{k+\mu_2+1/2}{k+\mu_1+1/2} \right]\quad &\text{for odd powers of }z.
\end{cases}
\end{align}
The remaining mismatched cases of $\Psi$ are simply given by $\Psi = [1+i]$.

The disentangling procedure rests on the fact that an orthogonal coordinate system of the complex plane can be devised such that one of the components of the vectors in these coordinates is independent of the mismatched terms. More precisely, rotating the complex plane under the multiplication by $e^{i\pi/4}$, one maps the matching terms to some vectors on the unit circle and the remaining ones are purely imaginary. Taking the real part of the result, we obtain an expression that only involves one degree of the Bannai-Ito polynomials per power of $z$. Let us now calculate the change in the normalization of each asymptotic function that this procedure induces. The rotation in the complex plane leads to
\begin{align*}
e^{i \frac{\pi}{4}}: \Psi_U \mapsto \Psi_U' = i \sqrt{2},\quad \Psi_A \mapsto \Psi_A' = e^{i \frac{\pi}{4}}\Psi_A.
\end{align*}
Taking the real part, the desired terms remain whereas the undesired ones vanish, leading to the required disentanglement. The real part of the rotated $\Psi_A$ is
\begin{align*}
\re{e^{i\frac{\pi}{4}}\Psi_A} = 
\begin{cases}
\displaystyle \frac{1}{\sqrt{2}}\left(1+\frac{k}{k+\mu_1+\mu_2}\right) &\text{for even powers of }z,\\
\\
\displaystyle \frac{-1}{\sqrt{2}}\left(1+\frac{k+\mu_2+1/2}{k+\mu_1+1/2}\right) &\text{for odd powers of }z.
\end{cases}
\end{align*}
Using the above, the transformed asymptotic wavefunctions, written $\widetilde{\mathcal{Y}}_K^N(z)$, are then monomials in $z$
\begin{align}
\widetilde{\mathcal{Y}}_K^N(z) = C_{K,N}^{\mu_{1},\mu_{2},\mu_{3} }\, z^{K},
\end{align}
where the coefficients are as follows, for $K=2k+p$ and $N=2n+t$ with $p,t \in \{ 0,1 \}$,
\begin{multline}\label{coeffMonomials}
C_{K,N}^{\mu_{1},\mu_{2},\mu_{3}} = \frac{(-1)^{p}}{2 \, \sqrt{k! (n-k+pt-p)!} } \Bigg[\frac{\Gamma(n+k+\mu_1+\mu_2+1 + p + t - pt)}{\Gamma(k+\mu_1+1/2+p)\Gamma(k+\mu_2+1/2+p)} \\
\times \frac{\Gamma(n+k+\mu_1+\mu_2+\mu_3+3/2+pt)}{\Gamma(n-k+\mu_3+1/2+t(1-p)) \Gamma(k+\mu_1+\mu_2+1)} \Bigg]^{1/2}.
\end{multline}

Acting with the same transformation on \eqref{fullGenFunEven} leads to the proper generating function. Writing $S=2s+p\in\{0,...,N\}$ with $p\in\{0,1\}$, for $N=2n+t \in \mathbb{N}$ with $t\in\{0,1\}$, one arrives at

\begin{multline}\label{properGenFunEven}
   \widetilde{\mathcal{Z}}_S^N(z) = \sum\limits_{u=0,1} z^{t+u}(1-z^2)^s \Big[(1-z^2)^{p-u} \\
   \times\, U_{S}^{u}\, P_{n-s-p}^{(2s+2p+\mu_2+\mu_3,\mu_1-1/2+t)}(2z^2-1) P_{s+p-u}^{(\mu_3-1/2+u,\mu_2-1/2+u)}\Big(\frac{\scriptstyle{z^2+1}}{\scriptstyle{z^2-1}}\Big) \\
   - z\, L_{S}^{u}\, P_{n+t-s-1}^{(2s+1+\mu_2+\mu_3,\mu_1+1/2-t)}(2z^2-1) P_{s}^{(\mu_3+1/2-u,\mu_2-1/2+u)}\Big(\frac{\scriptstyle{z^2+1}}{\scriptstyle{z^2-1}}\Big) \Big].
\end{multline}
where
\begin{align*}
 U_{S}^u = \frac{(-1)^{pu}}{\sqrt{2}} A_{S} B_{S} E_{S}^{1-2u}\xi_{S}^{+}, \qquad L_{S}^u = \frac{(-1)^{u(p+1)}}{\sqrt{2}} A_{S} B_{S}^{-1} F_{S}^{1-2u}\xi_{S}^{-}.
\end{align*}
The proper generating function decomposition is then expressed as
\begin{equation}
\widetilde{\mathcal{Z}}_S^N(z) = \sum\limits_{K=0}^N \, \mathcal{R}_{S,K,N}^{\mu_1,\mu_2,\mu_3}\, \widetilde{\mathcal{Y}}_K^N(z) = \sum\limits_{K=0}^N \, \mathcal{R}_{S,K,N}^{\mu_1,\mu_2,\mu_3} C_{K,N}^{\mu_{1},\mu_{2},\mu_{3} }\, z^{K},
\end{equation}
with $C_{K,N}^{\mu_{1},\mu_{2},\mu_{3} }$ as in \eqref{coeffMonomials} and where the Racah coefficients $\mathcal{R}_{S,K,N}^{\mu_1,\mu_2,\mu_3}$ as given in \eqref{RacahWave} are proportional to the Bannai-Ito polynomials.

\section{Conclusion}
We have derived generating functions for the Bannai-Ito orthogonal polynomials by exploiting the fact that these polynomials present themselves as the Racah coefficients for the $\mathfrak{osp}(1|2)$ Lie superalgebra. This derivation was done using an appropriate asymptotic expansion of Dunkl oscillators wavefunctions.

As the Bannai-Ito polynomials can be obtained as a $q\to -1$ limit of the Askey-Wilson polynomials or q-Racah polynomials, one could ask if their generating functions admit as limits the generating function derived here. However, all of the possible generating functions for these polynomials have limits with different truncation conditions for the Bannai-Ito polynomials than the ones used in this paper and do not correspond to to the result we derived.

This generating function for the Bannai-Ito polynomials might have interesting combinatorial interpretations \cite{Bannai1984}. Various orthogonal polynomials are obtained as limits of the Bannai-Ito polynomials. It would be interesting to investigate how generating functions for these polynomials can be recovered from the one obtained here.

\section{Aknowledgements}
The authors would like to thank Vincent Genest for inspiring discussions. The research of Geoffroy Bergeron is supported by scholarships of the Natural Science and Engineering Research Council of Canada (NSERC) and of the Fond de Recherche du Qu\'ebec - Nature et Technologies (FRQNT). The research of Luc Vinet is supported in part by a Discovery Grant from NSERC.

\bibliographystyle{amsplain}
\bibliography{Bannai-Ito.bib}

\end{document}